\documentclass{PoS}

\title{Hemisphere Mixing: a Fully Data-Driven Model of QCD Multijet Backgrounds for LHC Searches}

\ShortTitle{Hemisphere Mixing: a Fully Data-Driven Model of QCD Multijet Backgrounds for LHC Searches}

\author{P. de Castro Manzano, M. Dall'Osso, \speaker{T. Dorigo}\\
          INFN Padova\\
          E-mail: \email {pablo.decastromanzano@pd.infn.it},
                  \email{dorigo@pd.infn.it},
                  \email {martino.dallosso@pd.infn.it}}

\author { L. Finos, G. Kotkowski, G. Menardi, B. Scarpa\\
       Padova University\\
       E-mail: \email{livio.finos@unipd.it},
       \email {polandgreg@gmail.com},
       \email {menardi@stat.unipd.it},
       \email {scarpa@stat.unipd.it}}

\abstract{A novel method is proposed here to precisely model the multi-dimensional features of QCD multi-jet events in hadron collisions. The method relies on the schematization of high-$p_T$ QCD processes as  $2 \to 2$ reactions made complex by sub-leading effects. The construction of libraries of hemispheres from experimental data and the definition of a suitable nearest-neighbor-based association map allow for the generation of artificial events that reproduce with surprising accuracy the kinematics of the QCD component of original data, while remaining insensitive to small signal contaminations. The method is succinctly described and its performance is tested in the case of the search for the $\rm{hh \to b \bar b b \bar b}$ process at the LHC.}

\FullConference{The European Physical Society Conference on High Energy Physics\\
                 5-12 July\\
                 Venice, Italy}

\usepackage{graphicx}
\begin{document}

\noindent
Quantum Chromo-Dynamical (QCD) processes yielding multi-jet final states often constitute a problematic background to searches for rare phenomena in hadron-hadron collisions. The most advanced Monte Carlo (MC) generators can nowadays be trusted to produce a reliable modeling of the final state under investigation. Yet the huge cross sections of the involved processes call for prohibitively large simulations; limitations in the available computing power then prevent the use of QCD MC samples for modeling purposes, or affect the statistical accuracy of the resulting measurements.

% check BR(Hbb)

One example of the above situation are Large Hadron Collider (LHC) searches for pair-produced Higgs bosons in the four-b-quark final state, $\rm{hh \to b \bar{b} b \bar{b}}$. The study of Higgs boson pair production is the most straightforward way to determine the Higgs self-coupling $\lambda$, a parameter that may distinguish the Standard Model from many new physics scenarios involving anomalous couplings of the Higgs field\cite{anomalhiggs}. The large branching fraction of $\rm{h \to b \bar b}$ decay makes events with four b-quark jets attractive for that purpose. Unfortunately, the rate of QCD processes yielding the same final state dwarfs the rare $hh$ decay signal. The extraction of a cross section measurement in that channel thus demands a very precise multi-dimensional modeling of the observable features of QCD processes, enabling background rejection by multi-variate classifiers and an estimate of the signal component in selected data.

We describe here a novel experimental technique, called "hemisphere mixing", designed to model QCD multi-jet processes in LHC collisions, and we illustrate the statistical properties of the resulting model in the case of $\rm{hh \to b \bar b b \bar b}$ searches.  Event mixing techniques are not a novelty in experimental particle physics: they are frequently used in the study of symmetric electron-positron collisions, where the initial and the final state are quite clean, and the physics of the interaction makes the event simple to interpret. Conversely, in hadron collisions the complexity of the physics makes the application of mixing techniques much less straightforward, although examples do exist~\cite{Zjet,pbpb,angcor, BEC1, BEC2}. In the mentioned cases the elements subjected to mixing are individual particles, while here we use hadronic jets, as we are interested in the event features at the level of granularity where jets are the elemental observations. As jets are more direct messengers of the subnuclear reactions than individual hadrons, the creation of artificial events by the mixing of jet collections requires additional care. 

Multi-jet events can be described as the result of a simple tree-level $2 \to 2$ parton-parton scattering, made complex by ``second order'' effects such as QCD initial- and final-state radiation, pile-up, or multiple parton scattering. The kinematics of the two leading final-state partons, if properly estimated, may be used to identify events of similar characteristics. A mixing procedure exploiting that similarity may thus focus on the modeling of second-order effects. Artificial replicas of the original events can be created by mixing and matching subsets of the jets observed in different events. We explain how that is done in the following.

Let us consider a dataset of QCD multi-jet events, {\em e.g.} one collected by a suitable trigger in real LHC collisions, or simulated by a MC program. For each event an axis may be constructed on the plane transverse to the beams, the ``transverse thrust axis'', defined as the azimuthal angle $\phi_T$ (conventionally defined in $[0, \pi[$) which maximizes the transverse thrust quantity $T$: \par 

\begin{equation}
 T   =  \sum_j p_{T,j} |\cos (\phi_j-\phi_T)| 
\end{equation}

\noindent 
Once $\phi_T$ is known we may also define the related variable  $T_a$: \par

\begin{equation}
 T_a =  \sum_j p_{T,j} |\sin (\phi_j - \phi_T)|
\end{equation}

\noindent 
Above,  $j$ sums run on the collection of jets, and $p_{T,j}$ indicates the transverse momentum of jet $j$. The transverse thrust axis defines a plane orthogonal to it which divides the event in two separate hemispheres, each corresponding to a collection of the jets, with $\cos(\phi_j - \phi_T) >0$ or $\le 0$. The use of transverse coordinates in the construction permits to avoid dealing with the unknown boost of the center of mass of the hadron collision.

We characterize each hemisphere by its  number of jets ($N_j$), its number of b-tagged jets ($N_b$)\footnote{b-tagged jets are ones which are identified, based on their observed characteristics, as probable products of the hadronization of a b-quark.}, the sum of the projections of jet $p_T$ along the thrust axis ($T$),  the combined mass of the jets ($M$), the variable called $T_a$ (see Eq. 2), and the sum of the jets $p_z$ components, $P_z$. We may label them as $h_1(N_j, N_b, T, M, T_a, P_z)$ and $h_2(N_j, N_b, T, M, T_a, P_z)$. If we start from $N$ events, we obtain a library of $2N$ hemispheres. Using the hemisphere library we may create artificial replicas of the original events by pairing hemispheres observed in different events of similar kinematics. For each original event, composed of hemispheres $h_1$ and $h_2$, we look in the library for the two hemispheres $h_p$ and $h_q$ that are the {\em most similar} to $h_1$ and $h_2$, in the sense that they have the exact same value of $N_j$ and $N_b$, and have the {\em smallest distances} $D(1,p)$, $D(2,q)$ in the space spanned by the additional continuous variables $T$, $M$, $T_a$, $P_z$: \par

\small
\begin{equation}
D(1,p)^2 = \frac{(T(h_1)-T(h_p))^2}{V_T} + \frac{(M(h_1)-M(h_p))^2}{V_M} + 
           \frac{(T_a(h_1)-T_a(h_p))^2}{V_{T_a}} + \frac{(|P_z(h_1)|-|P_z(h_p)|)^2}{V_{P_z}}
\end{equation}
\begin{equation}
D(2,q)^2 = \frac{(T(h_2)-T(h_q))^2}{V_T} + \frac{(M(h_2)-M(h_q))^2}{V_M} +
           \frac{(T_a(h_2)-T_a(h_q))^2}{V_{T_a}} + \frac{(|P_z(h_2)|-|P_z(h_q)|)^2}{V_{P_z}}
\end{equation}
\normalsize

\noindent Above, denominators contain the variances $V_T$, $V_M$, $V_{T_a}$ and $V_{P_z}$ of the considered variables.The identified pair $h_p$, $h_q$ constitutes an entirely new event, as we prevent $h_p$ and $h_q$ from being equal to $h_1$ and $h_2$. We match the $P_z$ variables by considering only their absolute value (thus assuming, as it is safe to do for ATLAS and CMS, that detector acceptance to jets is forward-backward symmetric), and invert the sign of jet $p_z$ components in one of the two matched hemispheres if $sgn[P_z(h_1) P_z(h_2)] \neq sgn[P_z(h_p) P_z(h_q)]$. \, $h_p$ and $h_q$ are finally rotated along the azimuthal direction to match the original thrust axis of the modeled event. The procedure outlined above allows us to create an artificial dataset, of numerosity equal to that of the original data sample. Artificial events  are composed by hemispheres observed in original data, combined in a way that preserves their general features. Multi-dimensional statistical tests using a complete set of kinematic variables prove that the model reproduces the features of the original dataset quite accurately, if the latter is indeed constituted predominantly by QCD multi-jet events. When the original dataset contains a small fraction (say, a few percent) of events originated by a heavy particle decay, or Higgs pair production events, the mixing procedure {\em smears out} the features of the minority component, making them more similar to those of QCD: artificial data thus remain a robust model of the QCD component alone. This happens because the probability that the mixing procedure models a rare signal event with the combination of two hemispheres also originally belonging to signal events scales with the square of the signal fraction. That probability also depends on how ``recognizable'' are the signal hemispheres, according to the metric defined above. As long as $D$ uses variables that do not discriminate too strongly the signal from QCD multi-jet production, a small signal contamination  will not affect the results of the mixing procedure, which will still produce an artificial sample faithfully modeling the dominant process. This property makes the method quite attractive to model the large QCD background in small-signal searches.

\begin{figure}[h]
	\centering
	\includegraphics[width=12cm]{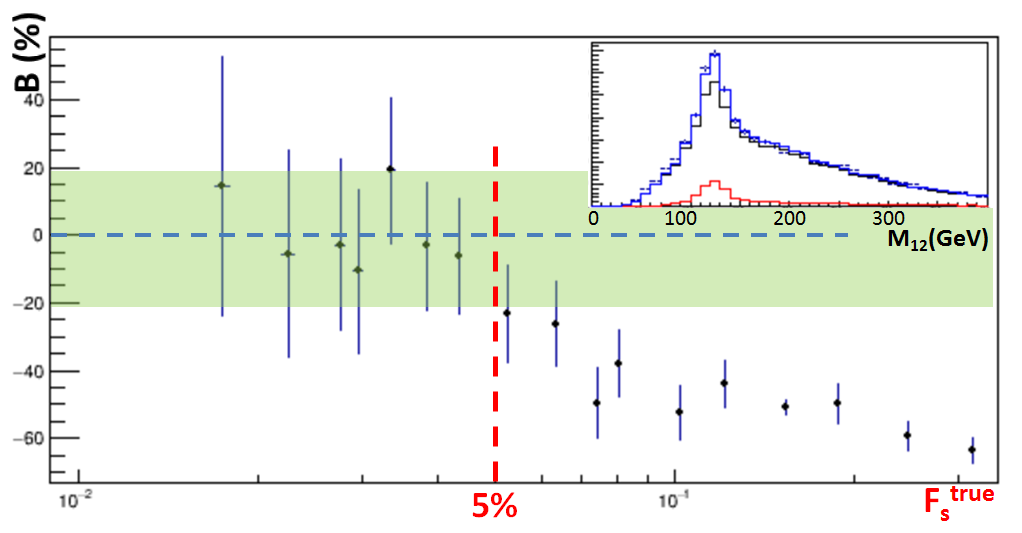}		
	\caption{ The percentage bias $B$ (black points) on the estimated signal fraction extracted from fits to the reconstructed Higgs boson mass distributions is drawn as a function of the true signal fraction $F_s^{true}$. The green band shows the level of bias considered acceptable for searches of new signals in hadron collider data.  The upper right inset shows the distribution of the leading jet pair mass $M_{12}$  of the QCD (black) and signal (red) components, their sum (blue), and the fit result (points with uncertainty bars) for a 5\% signal contamination.}
\end{figure}

The validity of the model is tested using fast-simulated LHC 13-TeV proton-proton collisions produced by the Delphes MC~\cite{delphes}. We consider a dataset corresponding to 5 inverse femtobarns of QCD multi-jet events and a detector simulation mimicking the average characteristics of ATLAS and CMS. We add to it different fractions of simulated  $\rm{hh \to b \bar b b \bar b}$ events in turn, creating several datasets of different composition. 

The datasets are reduced by the following event  selection. Jets are considered if they have $p_T>30$ GeV and pseudo-rapidity\footnote{Pseudo-rapidity is defined as $\eta = -\log(\tan \theta/2)$, where $\theta$ is the angle that a jet makes with the beams direction.} $|\eta|<2.5$, and events are kept if they contain  at least four b-tagged jets, $N_b \ge 4$. In order to reconstruct a variable sensitive to the signal component, we pair up the four b-tagged jets\footnote {In order to carry out the procedure described in the text, we order jets by the b-tagging discriminator variable~\cite{csv} and consider the leading four jets in the list.} such that the two jet pairs have minimum invariant mass difference; this defines two-jet masses $M_{12}$ and $M_{34}$ which exhibit peaking distributions at about $125$ GeV in signal events, and smoother distributions in QCD events. We finally extract an estimate of the signal fraction in each mixture dataset by a two-component likelihood fit to the two-dimensional distribution of $M_{12}$ and $M_{34}$, by using artificial data as a model of the QCD background alone, and the Higgs pair simulation as a model of the signal process. A comparison of the fitted with the true signal fraction allows us to draw some conclusions on the accuracy of the background model.

Figure 1 shows the results of the above procedure. The fit returns a unbiased estimate of the signal component if the latter is comparably small. For signal fractions above 5\% the signal contamination is seen to affect the corresponding artificial sample, which is used as a background-only model by the fit. It is generally agreed that biases up to 20\% in the estimated signal component can be considered acceptable in typical searches for new particles. Therefore we conclude that the modeling method is valid when small signals are sought, and it indeed is quite powerful, as it is entirely data-driven and the statistical precision of the model matches the one of real data.

The hemisphere mixing technique was used in a preliminary search for Higgs pair production in 2015 data by the CMS collaboration~\cite{cms15}. Its application to the larger 2016 dataset is under way.

\section* {Acknowledgements}

\noindent This report is part of a project that has received funding from the European Union's Horizon 2020 research and innovation programme under grant agreement No. 675440. M. Dall'Osso is supported by University of Padova grant CPDR155582.


\begin{thebibliography}{99}

\bibitem{anomalhiggs} V. Barger, L.L. Everett, C.B. Jackson, G. Shaugnessy, {\em Higgs-Pair Production and Measurements of the Triscalar Couplings at LHC (8,14)}, \emph{Phys. Lett.} {\bf B 728} (2014) 433, DOI: 10.1016/j.physletb.2013.12.013.

%\bibitem{anomalhiggs2} M. Slawinska, W. van den Wollenberg, B. van Eijk and S. Bentvelsen, {\em Phenomenology of the trilinear Higgs coupling at proton-proton colliders}, arXiv:1408.5010 (2014).

\bibitem{Zjet} A.M. Sirunyan et al. (CMS Collaboration), {\em Study of Jet Quenching with Z+jet Correlations in Pb-Pb and pp Collisions at $\sqrt{s_{NN}}$=5.02 TeV},
Phys. Rev. Lett. 119, 082301, DOI: https://doi.org/10.1103/PhysRevLett.119.082301 .
%@techreport{CMS-PAS-HIN-15-013,
 %     title         = "{Study of Z+jet correlations in PbPb and pp collisions at
 %                      $\sqrt{s_{\rm NN}} = 5.02~\mathrm{TeV}$}",
 %     institution   = "CERN",
 %     author = "{The CMS Collaboration}",
 %     address       = "Geneva",
 %     number        = "CMS-PAS-HIN-15-013",
 %     year          = "2016",
 %     reportNumber  = "CMS-PAS-HIN-15-013",
 %     url           = "{https://cds.cern.ch/record/2156179}",
%}
\bibitem{pbpb} S. Khachatryan et al. (CMS Collaboration), {\em Correlations between jets and charged particles in PbPb and pp collisions at  $\sqrt{s_{NN}}=2.76$ TeV}, \emph{Journ. High En. Phys.} {\bf 2} (2016), 156, DOI: 	10.1007/JHEP02(2016)156.

%@Article{Khachatryan2016,
%author="{The CMS Collaboration}",
%title="{Correlations between jets and charged particles in PbPb and pp collisions at                                                                                                                                                       s                                                  N                          N                                                                                      =                    2.76                                                  {\$}{\$} {\backslash}sqrt{\{}s{\_}{\{}{\backslash}mathrm{\{}NN{\}}{\}}{\}}=2.76 {\$}{\$}               TeV}",
%journal="Journal of High Energy Physics",
%year="2016",
%month="Feb",
%day="23",
%volume="2016",
%number="2",
%pages="156",
%issn="1029-8479",
%doi="10.1007/JHEP02(2016)156",
%url="https://doi.org/10.1007/JHEP02(2016)156"
%}

\bibitem{angcor} S. Khachatryan et al. (CMS Collaboration), {\em Observation of long-range, near-side angular correlations in proton-proton collisions at the LHC},  \emph{Journ. High En. Phys.} {\bf 9} (2010) 91, DOI: 	10.1007/JHEP09(2010)091.
%@Article{Khachatryan2010,
%author="{The CMS Collaboration}",
%title="{Observation of long-range, near-side angular correlations in proton-proton collisions at the LHC}",
%journal="Journal of High Energy Physics",
%year="2010",
%month="Sep",
%day="27",
%volume="2010",
%number="9",
%pages="91",
%issn="1029-8479",
%doi="10.1007/JHEP09(2010)091",
%url="https://doi.org/10.1007/JHEP09(2010)091"
%}

\bibitem{BEC1} S. Khatrchyan et al. (CMS Collaboration), {\em Measurement of Bose-Einstein correlations in pp collisions at root s=0.9 and 7 TeV}, \emph{Journ. High En. Phys.} {\bf 05} (2011) 29, DOI: 10.1007/JHEP05(2011)029. 

\bibitem{BEC2} S. Khatrchyan et al. (CMS Collaboration), {\em First Measurement of Bose-Einstein Correlations in Proton-Proton Collisions at root s=0.9 and 2.36 TeV at the LHC}, {\em Phys. Rev. Lett.} {\bf 105} (2010) 032001, DOI:  http://hdl.handle.net/1721.1/60901.

\bibitem{delphes} J. de Favereau et al., {\em DELPHES 3, A modular framework for fast simulation of a generic collider experiment}, \emph{Journ. High En. Phys.} {\bf 02} (2014) 57, DOI: 10.1007/JHEP02(2014)057.

\bibitem{csv} CMS Collaboration, {\em Identification of b-quark jets with the CMS experiment} {\emph J. Instrum.}  8 (2013) 04013, Doi: 10.1088/1748-0221/8/04/P04013.

%\bibitem{higgssearchprocedure} S. Chatrchyan et al. (CMS Collaboration), {\em Combined results of searches for the standard model Higgs boson in pp collisions at sqrt(s)=7 TeV}, \emph{Phys. Lett.} {\bf B710} (2012) 26.

\bibitem {cms15} CMS Collaboration, {\em Search for non-resonant pair production of Higgs bosons in the $b \bar b b \bar b$ final state with 13 TeV CMS data}, CMS-PAS-HIG-16-026 (2016), https://cds.cern.ch/record/2209572.

\end{thebibliography}
\end{document}